\documentclass[12pt,preprint]{aastex}

\lefthead{}
\righthead{Argon Abundances in Orion Association B Stars}

\begin{document}

\title{Argon Abundances in the Solar Neighborhood: \\
        Non-LTE Analysis of Orion Association B-type Stars\altaffilmark{1}}

\author{Thierry Lanz}
\affil{Department of Astronomy, University of Maryland, College Park MD 20742, USA;
lanz@astro.umd.edu}

\author{Katia Cunha\altaffilmark{2}}
\affil{National Optical Astronomy Observatory,
Casilla 604, La Serena Chile; kcunha@noao.edu}

\author{Jon Holtzman}
\affil{New Mexico State University, Las Cruces NM 88003, USA;
holtz@nmsu.edu}

\author{Ivan Hubeny}
\affil{Steward Observatory, University of Arizona, Tucson AZ 85712, USA;
hubeny@aegis.as.arizona.edu}

\begin{abstract}
Argon abundances have been derived for a sample of B main-sequence stars in the Orion association.
The abundance calculations are based on NLTE metal line-blanketed model atmospheres calculated
with the NLTE code TLUSTY and an updated and complete argon model atom. 
We derive an average argon abundance for this young population of
A(Ar)~=~6.66 $\pm$~0.06. While our result is in excellent agreement with a recent analysis
of the Orion nebula, it is significantly higher than the currently recommended solar
value which is based on abundance measurements in the solar corona. Moreover, the derived argon
abundances in the Orion B stars agree very well with a measurement from a solar impulsive flare during which
unmodified solar photospheric material was brought to flare conditions. We therefore
argue that the argon abundances obtained independently for both the Orion B stars and the Orion nebula
are representative of the disk abundance value in the solar neighborhood. The lower
coronal abundance may reflect a depletion related to the FIP effect. We propose
a new reference value for the abundance of argon in the solar neighborhood,
A(Ar)~=~6.63 $\pm$~0.10, corresponding to Ar/O~=~0.009 $\pm$~0.002.

%We propose
%new reference values for the abundance of neon and argon in the solar neighborhood,
%A(Ne)~=~8.09 $\pm$~0.06, and A(Ar)~=~6.63 $\pm$~0.10, corresponding to Ne/O~=~0.26 and
%Ar/O~=~0.009.

%Thierry, I am not sure if we should  keep this info about the neon abundance in the abstract because
%we are not really dealing with neon in this paper. I think that having this
%in the conclusions would be good but not in the abstract (Just a suggestion!
%however, happy if prefer to.

\end{abstract}


\altaffiltext{1}{Based on observations obtained with the Apache Point Observatory
3.5-meter telescope, which is owned and operated by the Astrophysical
Research Consortium.}

\altaffiltext{2}{On leave from Observat\'orio Nacional - MCT; Rio de Janeiro, Brazil}

\setcounter{footnote}{2}

% -----------------------------------------------------------------------------

\section{INTRODUCTION}

Establishing the abundance of noble gases in the Sun has been a notoriously
complex task, with significant uncertainties still remaining in some cases.
These difficulties primarily ensue from the atomic structure of noble gases, 
in particular their high first ionization potential and the large excitation energies
of the excited levels in neutral atoms. Noble gases are therefore in the neutral
ground state to a great extent in the solar photosphere and, consequently,
no spectral lines of these elements have been identified in the photospheric
solar spectrum. Well-established model stellar atmosphere and quantitative
spectroscopic techniques therefore cannot be applied to study noble gases in
the Sun. Furthermore, these species are volatile. Meteoric studies thus yield
only a lower limit of the actual abundance of noble gases in the solar system and,
therefore, are of limited usefulness for inferring the solar argon abundance.

While the solar helium abundance can be determined with a great degree of accuracy
from helioseismology, abundance studies of heavier noble gases depend on
the analysis of coronal lines and {\sl in-situ\/} measurements of energetic particles
from the solar wind. These techniques have greater uncertainty than photospheric
abundance studies which have been applied to many other chemical elements.
Moreover, measurements are made relative to a reference element, and
they actually yield abundance ratios such as the Ar/O ratio.
In a recent compilation of solar abundance works, \cite{AGS05} recommend a solar
argon abundance, A(Ar)$_{\odot} = 6.18$.\footnote{Abundances are quoted in number densities
throughout the paper, in the standard scale where A(H)~=~12.0.} This value
significantly revises downward the earlier standard value
(A(Ar)$_{\odot} = 6.55$, \cite{SUN89}; A(Ar)$_{\odot} = 6.50$, \cite{lodders08}),
mostly because of the downward revision of the solar oxygen abundance
\citep{allende01, asplund04}.

Because of the shortcomings of solar studies, it appears that abundance measurements
in the solar neighborhood may shed more light on the argon abundance at near-solar metallicity.
Indeed, this approach is suggested by the similar case of neon, for which
recent analyses in B-type stars \citep{cunha06},
in the Orion nebula \citep{esteban04}, and in the interstellar medium (ISM) toward
the Crab nebula \citep{kaastra07}, provide very consistent results, Ne/O~=~0.26$\pm$0.01,
and therefore indicate that the currently adopted solar value, (Ne/O)$_\odot$=0.15,
most likely underestimates the actual neon abundance in the solar photosphere and in the solar
neighborhood. Defining a reliable set of standard chemical
abundances is an essential undertaking, because abundance studies are a common tool used
to study physical processes which are revealed by changes in the chemical composition 
of various astrophysical media, for instance including chemical anomalies in the
solar corona, in the ISM (e.g., depletion into grains, ISM ionization),
or in stellar photospheres (e.g., mixing of nucleosynthesis products).

The optical spectra of early-type stars reveal a number of unblended, weak
\ion{Ar}{2} lines. Despite appearing to be the most fitting targets from which to derive
the argon abundance in the solar neighborhood, very limited abundance work on argon
in early-type stars has been carried out to this day. 
%%%Trundle et al. (2001) analyzed argon, along with other elements, in a peculiar metal-rich star. 
%%%THierry, see if you think it is ok to add a reference Trundle et al (2001) work here or somewhere else?
%%%She sent us non-published eqws for Ar II lines from this paper which helped us in the beginning
%%%of our analysis. 
\cite{keenan90} analyzed two \ion{Ar}{2} lines in the blue spectrum of 5 bright B-type stars
in the field, assuming
Local Thermodynamic Equilibrium (LTE), and they derived an argon abundance, A(Ar)$ = 6.49\pm0.1$.
\cite{holmgren90} supplemented this study with an investigation of the importance of departures from LTE, 
which were found to be small. The NLTE result (A(Ar)$ = 6.50\pm0.05$) thus does not
differ significantly from the LTE analysis; these NLTE calculations, however, were
based upon non-blanketed model atmospheres. 

In this paper we report on the results of a new NLTE analysis of a sample of
10 main-sequence B-type stars which are members of the young Orion association.
The argon abundances are derived from NLTE line-blanketed model atmosphere
calculations. We argue that  
the argon abundances obtained for this young population 
may well represent the actual argon content of the solar neighborhood and that
the Orion B stars can be used as good proxies to define the argon abundance in the
solar photosphere.
Sect.~2 briefly presents the new observational material,
while the NLTE calculations are detailed in Sect.~3. We discuss our
results in Sect.~4, comparing them to other determinations of the argon abundance
in different sites. In our conclusions (\S5), we propose a new 
reference value for the argon abundance in the solar neighborhood.

% -----------------------------------------------------------------------------

\section{OBSERVATIONS AND DATA REDUCTION}

The sample under study consists of 10 main-sequence early B-type stars, which
have been previously analyzed in our recent work devoted to neon \citep{cunha06}.
New spectra were obtained on the nights of 2007 January 29/30, February
5/6, and March 2/3, using the ARC echelle spectrograph on the 3.5~m
telescope at the Apache Point Observatory. The ARC echelle covers the
entire visible spectrum on a single 2048x2048 detector with a resolving power
of $R\approx$ 35,000. Exposure times for the different targets are listed in
Table~\ref{tab:exp}. Because of partly cloudy conditions, some stars were
observed repeatedly in order to achieve a signal-to-noise ratio of at least 100 in 
the co-added spectra.

\begin{figure}
\epsscale{.8}\plotone{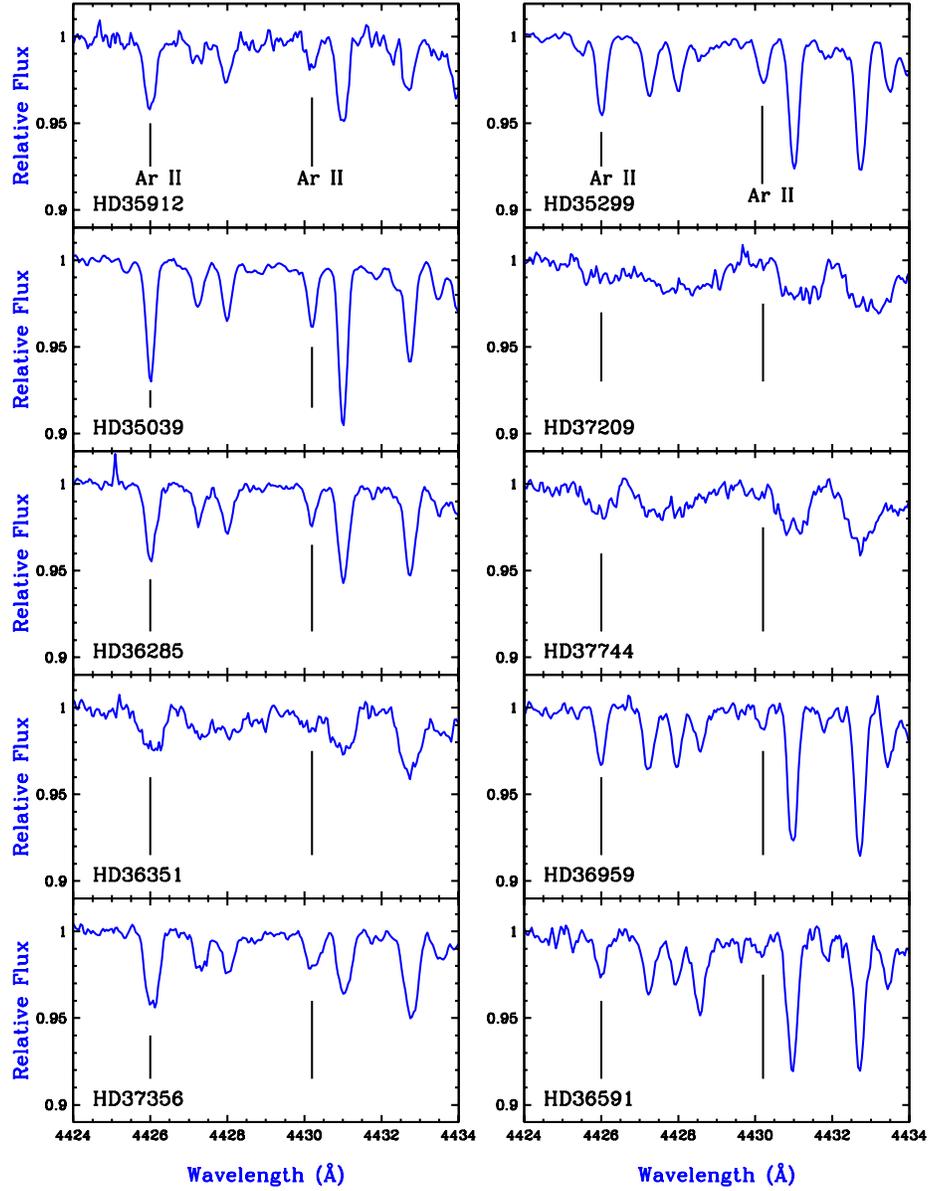}
\caption{\label{fig1} \ion{Ar}{2}$\lambda\lambda$4426, 4430 lines in the whole stellar sample.
The stars in the panels are arranged in order of increasing $T_{\rm eff}$. 
}
\end{figure}

The data were reduced and the spectra extracted following fairly standard
procedures. Bias frames and flat field exposures were combined; flat
frames included frames with a blue filter to insure a high signal-to-noise ratio
across the entire spectrum.
Individual orders were extracted from both the flats and
the object frames, after subtraction of a relatively small scattered light
component. Some care was taken with the extraction parameters because of
the relatively narrow width of the orders in the spectrograph. The
extracted spectra were flattened using the one dimensional extracted
flats. Wavelength calibration was performed with Th-Ar exposures,
and applied to the object spectra. Noise spectra were obtained
using the known gain and readout noise of the detector. The extracted spectra
were coadded when multiple observations were available. Finally,
the individual orders were combined, using the noise spectra to properly
weight the different orders in regions where there was spectral overlap.
Sample spectra of all target stars in the region around $\lambda$ 4429 \AA\
are shown in Fig.~\ref{fig1}.

% ------------------------------------------------------------------------------------

\section{NON-LTE ABUNDANCE CALCULATIONS}

The NLTE calculations have been performed with the NLTE model stellar atmosphere
code TLUSTY \citep{TLUSTY88, TLUSTY95}.\footnote{See 
http://nova.astro.umd.edu.} The TLUSTY model atmospheres assume
hydrostatic and radiative equilibria, and incorporate all essential sources of
opacity while allowing for departures from LTE. A large grid of NLTE fully
line-blanketed model atmospheres of early B-type stars has been made available
recently (BSTAR2006; Lanz~\& Hubeny 2007). The present calculations follow 
the BSTAR2006 models, but for larger Ne~{\sc i-ii}\  model atoms \citep{cunha06}
and for adding extensive Ar~{\sc i-iv}\  model atoms  (see \S\ref{modelAtom})
in the NLTE calculations. The NLTE argon calculations thus benefit from an
extensive and detailed NLTE treatment of all background opacity sources.

We have adopted for the new models the same stellar parameters of the target
stars as \cite{cunha06}, see Table~\ref{tab:resul}. The stellar parameters 
were derived by \cite{cunha92}, who provided a detailed discussion about
uncertainties. We examine their implications for the argon abundances
in \S\ref{abundances}. Like in BSTAR2006, a constant microturbulent velocity
of 2 km s$^{-1}$ was assumed in the NLTE model atmosphere calculations. 

\setcounter{footnote}{3}

NLTE synthetic spectra were then calculated with the spectrum synthesis code
SYNSPEC\footnotemark\  using the NLTE level populations for all levels
included in the TLUSTY models. We have adopted in this step a microturbulent velocity
of 5 km s$^{-1}$ that is typical for early B-type main-sequence stars. We
stress, however, that the \ion{Ar}{2} lines analyzed in all 10 stars
are relatively weak and, therefore, the results are quite insensitive
to the microturbulence parameter (see \S\ref{abundances}).

\subsection{Argon model atom}
\label{modelAtom}

The argon model atom consists of 71 levels of \ion{Ar}{1}, 54 levels of \ion{Ar}{2}, 
44 levels of \ion{Ar}{3}, 36 levels of \ion{Ar}{4}, plus the ground state of
\ion{Ar}{5}. Because the level structure of \ion{Ar}{1} is better described with
intermediate $jK$-coupling (similarly to \ion{Ne}{1}), the \ion{Ar}{1} model
atom was constructed to account for fine structure. It includes the lowest
71 fine-structure levels (up to the $6s$ and $6p$ levels,
$E < 121,500$\,cm$^{-1}$). Experimental energy level data and line oscillator
strengths were extracted from the Atomic and Spectroscopic Database at
NIST \citep{NIST}.\footnote{See http://physics.nist.gov/PhysRefData/ASD/index.html.}
The photoionization cross-sections were taken from the Opacity Project (OP) database
Topbase \citep{topbase}.\footnote{See http://vizier.u-strasbg.fr/topbase.}
Since OP calculations assume $LS$-coupling, we followed a procedure analogous
to the case of \ion{Ne}{1} \citep{seaton98} for relating fine-structure levels
to $LS$-multiplet levels and thus for assigning the appropriate photoionization
cross-section to each level. The collisional excitation rates were considered using
the \cite{vanrege62} formula, including a modification for neutral atoms (as also
used by \cite{auer73}); for collisional ionization the Seaton formula was used
(for a synopsis of expressions, see \cite{TLUSTY88}).

For the Ar~{\sc ii-iv}\  ions, we closely followed the approach described in
\cite{OS02} regarding the treatment and inclusion of atomic data in model atmosphere
calculations. In particular, the bulk of the data are taken from Topbase. The level
energies were updated with the more accurate experimental values whenever available
in the NIST ASD database. For a detailed description of the process of setting up
the atomic data, the reader should refer to \S4 of \cite{OS02}. In particular, we
include in each model ions all the energy levels which are in Topbase and below
the respective ionization limit, that is, typically up to $n=10$.
The \ion{Ar}{2} model atom includes individually the lowest 42 levels 
up to $4d~ ^2$D ($E < 193,000$\,cm$^{-1}$). The 206 higher levels are grouped
into 12 superlevels, 6 in the doublet and 6 in the quartet systems.
The \ion{Ar}{3} model atom includes the lowest 27 levels 
up to $3d~ ^3$P$^o$ ($E < 215,000$\,cm$^{-1}$). The 349 higher levels are grouped
into 17 superlevels, 6 in the singlet, 6 in the triplet and 5 in the quintet
systems. The \ion{Ar}{4} model atom includes the lowest 22 levels 
up to $4s~ ^2$D ($E < 270,000$\,cm$^{-1}$), and the first 3 sextet levels.
The 322 higher levels are grouped
into 10 superlevels, 5 in the doublet and 5 in the quartet systems. Five
excited sextet levels, below the ionization limit, are grouped into an
additional superlevel.

\subsection{Abundances from selected Argon lines}
\label{abundances}

The optical spectra of early B-type stars contain a number of unblended \ion{Ar}{2} lines.
The sample of 11 \ion{Ar}{2} lines analyzed in this study is presented in Table~\ref{tab:line}.
The \ion{Ar}{2} $gf$-values for these transitions were extracted from \cite{bennett65}
via the NIST ASD database.
The accuracy assessment of these oscillator strengths by NIST is very high, with
7 values estimated to be as accurate as 3\%\,  while the 4 other lines have
estimated uncertainties smaller than 10\%. The linelist used in computation of the synthetic spectra
around each of the \ion{Ar}{2} transitions is an update of the \cite{CD23} linelist and
is available from the TLUSTY website.\footnotemark[4]

\begin{figure}
\epsscale{.8}\plotone{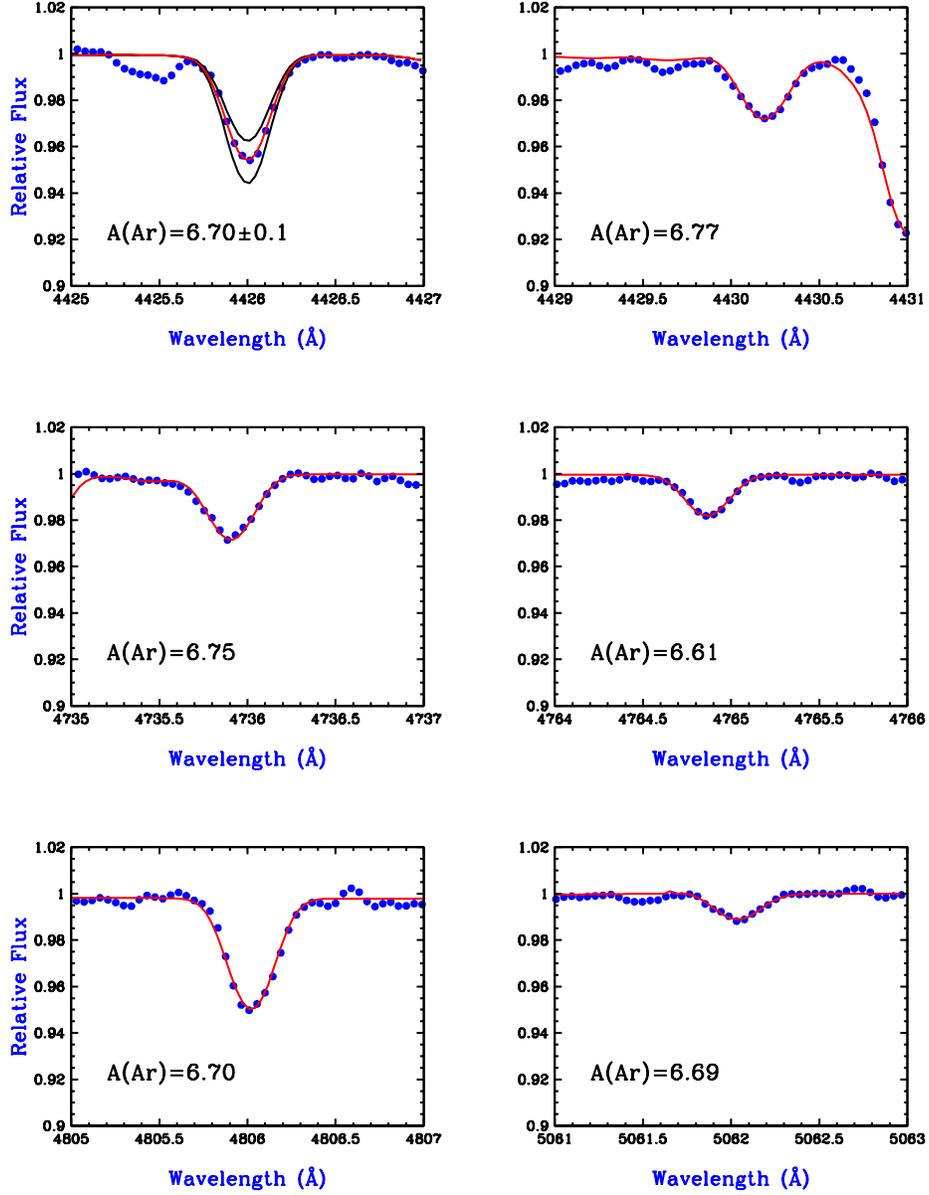}
\caption{\label{fig2} Sample \ion{Ar}{2} lines in the spectrum of the B1.5V star, HD 35299:
observed spectrum (blue points) and best-fit NLTE model spectrum (gray line;
red line in electronic edition). The effect of changing the argon abundance by 0.1~dex
on the predicted line profile is illustrated in the top left panel.
}
\end{figure}

Argon abundances were derived from best fits between synthetic and observed spectra. 
Sample fits for 6 \ion{Ar}{2} lines ($\lambda \lambda$ 4426; 4430; 4735; 4764; 4806 and 5062 \AA)
in the star HD35299 are shown in Fig.~\ref{fig2}. 
The best-fit synthetic profiles for each line 
were calculated with the argon abundances indicated in the different panels of the figure. 
The sensitivity of the fits to changes in the argon abundance by $\pm$ 0.1 dex is 
illustrated for the \ion{Ar}{2}$\lambda$4426 line in the top left panel of Fig.~\ref{fig2}. 
The final argon abundances for the target stars are given in Table~\ref{tab:resul}. They
are the averages of the abundances derived from the individual \ion{Ar}{2} lines. 
The corresponding standard deviations represent the line-to-line scatter, which may be
attributed in part to the remaining uncertainties on the $f-$values (up to 0.05~dex).
For completeness, we also show in Table~\ref{tab:resul} the oxygen abundances
obtained previously for the target stars, because coronal studies provide the abundance ratio Ar/O. 
The oxygen abundances are extracted from \cite{cunha94}; see \cite{cunha06} for a discussion of
consistency checks between these oxygen abundances and TLUSTY. Abundances are listed
as number densities in the standard logarithmic scale, where the hydrogen abundance is A(H)=12.

The uncertainties in the derived argon abundances can be evaluated by investigating abundance
changes resulting from model atmospheres computed with modified stellar parameters. The error
budget in the argon abundances is dominated by the uncertainties in $T_{\rm eff}$. The adopted
effective temperatures are derived from Str\"omgren photometry; \cite{cunha94} carefully assessed
the uncertainties to $\delta T_{\rm eff}$= $\pm$3$\%$ at most. We verified that the adopted 
$T_{\rm eff}$ values and uncertainties are consistent with temperature-sensitive spectroscopic
diagnostics such as the \ion{Si}{3}$\lambda\lambda$4553, 4568, 4575 lines. These uncertainties
translate to an error on the derived abundances, $\delta$A(Ar)= $\pm$0.08 dex. On the other hand,
the uncertainty on gravity, $\delta$log g= $\pm$0.1 dex, has a very limited effect on the abundances,
$\delta$A(Ar)= $\mp$0.01 dex.
%% The abundance
%% differences caused by changing the effective temperature by $\delta T_{\rm eff}$= $\pm$3$\%$ and the surface
%% gravity by $\delta$log g= $\pm$0.1 dex (typical uncertainties in the adopted stellar parameters; 
%% \cite{cunha92, cunha94}) correspond to $\delta$A(Ar)= $\pm$0.08 dex and $\mp$0.01 dex, respectively. 
The errors due to uncertainties in the adopted microturbulent velocities are small given that 
the sample \ion{Ar}{2} lines are all reasonably weak. A change in microturbulence of 1.5~km/s 
results in $\delta$A(Ar)= $\pm$0.01 dex. The total error from uncertainties in these stellar parameters
is $\pm$0.08 dex, using a quadratic error summation.

% -----------------------------------------------------------------------------

\section{DISCUSSION}

The argon abundance cannot be measured in the photospheres of the Sun and
of other cool stars. We discuss and compare argon abundances derived in
different sites, including the solar corona, {\sl in-situ\/} measurements in the
solar system, the interstellar medium (ISM), \ion{H}{2} regions and planetary
nebulae, along with our analysis of early-type stars.  The most
representative results are discussed in the next sections. This comparison provides
for a reliable determination of the abundance of argon in the solar
neighborhood, which can then serve as a reference point for further studies.
A similar overview of argon abundance determinations has been recently presented by
\cite{lodders08}.

\subsection{Argon in early-type stars and departures from LTE}

The weighted (as well as straight) average of the NLTE abundances measured
in 10 early B stars of the Orion association, and listed in Table~\ref{tab:resul},
is $<$A(Ar)$> = 6.66 \pm 0.06$. All the \ion{Ar}{2} lines give a consistent result,
as indicated by the line-to-line scatter for each star which is always smaller than
0.1~dex, and the abundances span a very tight range
($6.59\leq$ A(Ar) $\leq 6.76$). The error budget from uncertainties in  stellar parameters
is $\pm$0.08~dex. Furthermore, the available
$f$-values are of high accuracy, generally better than 3\%. We therefore 
believe that the accuracy of the derived argon abundance is likely of the order
of 0.1~dex or better.  

\cite{keenan90} carried out an early, but very limited, study of argon in 5 nearby field
B-type stars. They observed two \ion{Ar}{2} lines ($\lambda\lambda$4590, 4658).
One star was excluded from the sample because only \ion{Ar}{2}$\lambda$4590 was marginally
detected. Argon abundances were derived from equivalent widths and LTE model atmospheres.
Results from a second star (HR~1350) were also excluded because the derived abundance
differed by 0.25~dex relative to the 3 other stars, which was attributed to larger
uncertainties in the stellar parameters. They derived an abundance, A(Ar)$ = 6.49 \pm 0.1$,
based on 4 measurements only. Their value is 0.17~dex lower and barely
consistent with our results.
However, if we retain the abundance derived for HR~1350 that they excluded, and if we correct
for small systematic differences in the $f$-values (0.024 and 0.046~dex for
$\lambda\lambda$4590, 4658, respectively), we find a better agreement
(A(Ar)$ = 6.54 \pm 0.15$). The larger scatter seems to be a more likely estimate of
the actual uncertainties of their results.

\cite{holmgren90} supplemented Keenan et~al.'s study and evaluated the importance
of departures from LTE. They applied the NLTE line formation codes DETAIL and SURFACE
\citep{giddings81, butler84} and found that the corrections from LTE are
very small. Holmgren et~al. derived an argon abundance, A(Ar)$ = 6.50 \pm 0.05$. However,
their NLTE analysis is based on unblanketed NLTE model atmospheres, while Keenan et~al.
used metal line-blanketed LTE models. Their comparison therefore amalgamates two
different effects, namely direct NLTE effects in argon and differences arising from
the changes in atmospheric structure between blanketed and unblanketed models.

To ascertain better the extent of departures from LTE in argon, we have repeated our
analysis of the target star HD35299 using a Kurucz LTE blanketed model atmosphere and 
LTE spectrum synthesis. We can then make a direct comparison of two consistent analyses with and without
the assumption of LTE, but based on fully line-blanketed model atmospheres in both cases.
The LTE abundances obtained for the sample \ion{Ar}{2} lines are not significantly different
from the NLTE values: $<\delta$(NLTE - LTE)$>\approx$ 0.02 $\pm$ 0.01~dex, supporting
Holmgren et~al.'s earlier claim that departures from LTE are small.
LTE argon abundances can also be derived using the same NLTE model atmosphere, but only setting
argon populations to their LTE values. This second approach may isolate the non-LTE
effects in argon since the same model structure is adopted. In this case, our results
show that NLTE effects are also small, the NLTE abundances being larger than the LTE values
by $\approx$~0.03~dex. These results therefore support the earlier NLTE study of \cite{holmgren90}.
We stress however the importance of performing extensive NLTE calculations
when aiming at determining abundances with accuracies as high as 0.1~dex or better. Indeed,
even small corrections to LTE may easily introduce systematic errors of the order of
the expected accuracy.

%%% At physical conditions present in the line-forming region in early B star
%%% atmospheres, a large majority of argon is found in the Ar$^{2+}$ ground state.
%%% As already pointed out by \cite{keenan90}, small NLTE effects are therefore
%%% expected because the \ion{Ar}{2} optical lines are transitions between highly-excited
%%% Ar$^+$ levels that are strongly coupled by collisions to the Ar$^{2+}$ ground state. 

\begin{figure}
\epsscale{.9}\plotone{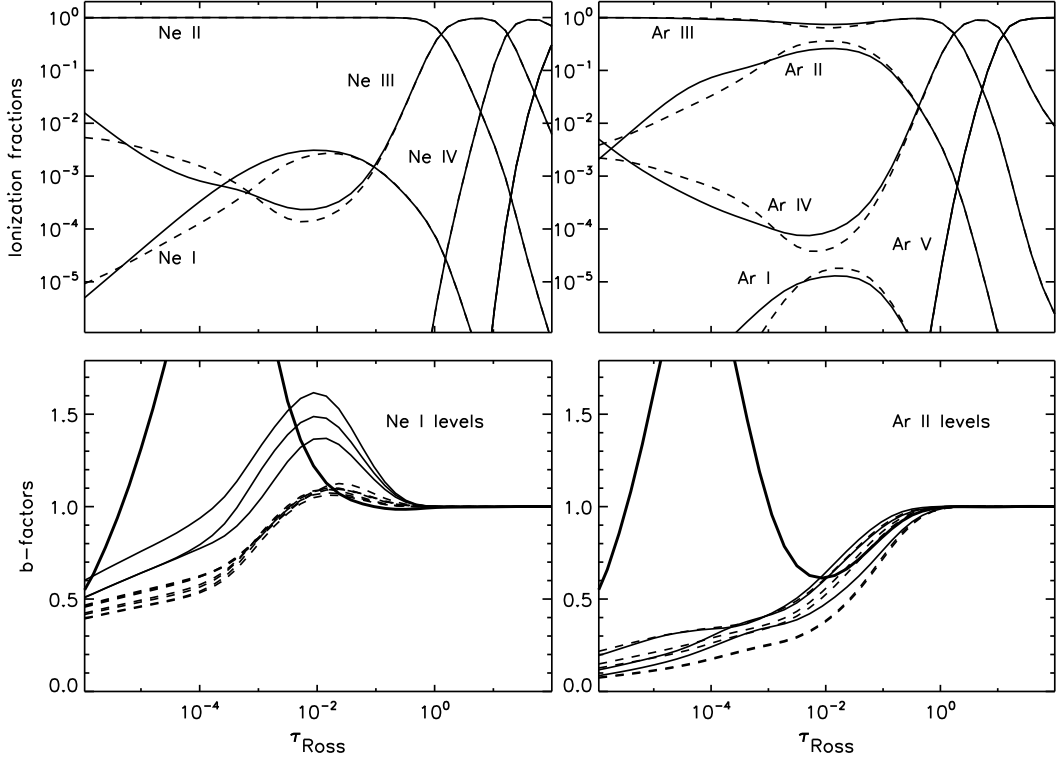}
\caption{\label{fig3} Top panels: NLTE ionization fractions of neon and argon as function
of depth in a model atmosphere, $T_{\rm eff} = 24,000$\,K, $\log g = 4.25$, and solar composition.
Dashed lines show LTE ionization fractions for the same physical conditions. Bottom panels: 
Departure coefficients of Ne and Ar$^+$ levels of the analyzed lines (lower levels: full lines;
upper levels: dashed lines); the thick lines show the departure coefficients of the Ne and Ar$^+$
ground states.
}
\end{figure}

We examine now the reasons why the departures from LTE are small for \ion{Ar}{2}
lines while quite significant for \ion{Ne}{1} lines \citep{sigut99, cunha06}. The
top panels of Fig.~\ref{fig3} show that the typical NLTE ionization shift toward higher
ionization is limited. Because of the high ionization potentials of neon and
argon ions, the ionization thresholds are found in the extreme ultraviolet where B stars
have a low flux. In this spectral range, the opacity is dominated by hydrogen and helium. The
bound-free transition from the ground states of neon and argon are optically thick and
in detailed radiative balance in the region of formation of neon and argon optical lines.
Consequently, the contribution of radiative photoionization to departures from LTE
remains small. Rather than by a global ionization shift, the line strengths are then better described by the
departure coefficients, $b_i = n_i / n_i^\star$, where $n_i$ is the population of level $i$
and $n_i^\star$ its LTE counterpart. The bottom panels of Fig.~\ref{fig3} show a very
different behavior for lines in the $3s - 3p$ \ion{Ne}{1} and in the $4s - 4p$ \ion{Ar}{2}
systems to which the analyzed lines belong. In both cases, the optical lines behave mostly
like lines in a 2-level atom because the resonance lines are optically thick at the depth
of formation of the optical lines. In the neon case, the (red) lines provide a downward channel,
overpopulating the lower levels, thus resulting in stronger absorption and yielding thus lower
abundances \citep{cunha06}. On the other hand, the Ar$^+$ levels are collisionally coupled
to the Ar$^{2+}$ ground state, as indicated by the fact that all Ar$^+$ levels show similar departure
coefficients. In first approximation, the ratio of the line source function to the Planck function
behaves like $S/B\propto b_u/b_l$. In the line forming region, our calculations typically give
$b_u/b_l\approx 0.98$, thus yielding only a small correction to LTE.
%%We attribute basically the different behavior of neon and argon levels to the
%%lower ionization potential of argon compared to neon, which results in stronger collisional
%%coupling to the dominant ion's ground state.
Finally, the \ion{Ne}{1} and \ion{Ar}{2}
resonance lines are formed in the outermost atmospheric layers,
around $\tau_{\rm Ross}\approx 10^{-4}$, where they provide a downward radiative channel,
leading to an overpopulation of the ground states and underpopulation of the excited
states.

Among hot stars, \cite{werner07} have recently reported the identification of
\ion{Ar}{7}~$\lambda$1063\  in very hot central stars of planetary nebulae and
(pre-)white dwarfs. Because of large remaining uncertainties on stellar parameters,
and because of processes such as nucleosynthesis and elemental diffusion occurring
in late evolutionary stages, their derived
argon abundances cannot be used as a reference value. Several stars
however show an argon abundance close to the value that we derived for B-type stars,
hence supporting the idea that there is no large production of argon in the
previous AGB evolution and that argon remains essentially unchanged in these stages.

\subsection{Nebular emission abundances}

\cite{esteban04} derived emission line abundances for several elements in the Orion nebula.
In particular, they obtained A(Ar)= 6.62 $\pm$ 0.05 when adopting their preferred value for
the temperature fluctuation parameter, $t^{2}$=0.02. The argon abundance in the Orion nebula
derived earlier by \cite{peimbert77} also ranges between A(Ar)=6.6 (with $t^{2}$=0)
and A(Ar)=6.7 (for $t^{2}$=0.35). There is therefore a good indication of a consensus
regarding the nebular argon abundance in Orion. Furthermore, the nebular analysis is
fully consistent with our results from B stars in the Orion association (A(Ar)= 6.66 $\pm$ 0.06),
even with the tight uncertainties of the two studies.

Argon abundances can also be derived from planetary nebulae (PNe). \cite{stanghellini06}
present argon abundances for a large sample of galactic PNe
and find an average argon abundance for the round nebulae (representing the lower mass
and probably local population) of $<$A(Ar)$>$~=~6.11. These argon abundances rely on optical 
argon lines and are deemed as more uncertain in their study. 
\cite{pottasch06} give
a summary of PN abundances from measurements recorded with the {\sl Infrared Space
Observatory\/}. The Galactic chemical gradient is apparent from their dataset
when plotting abundances as a function of galactocentric distances. Therefore, for comparison
with the Orion results, we have retained
only the PNe which are within 1~kpc from the Sun. The average argon abundance of the 7 PNe
satisfying this criterion is A(Ar) = 6.50$^{+0.14}_{-0.22}$. The average value is 0.12~--~0.16~dex
lower than the results from Orion. However, the scatter is much larger and we cannot conclude
that there is a significant difference. The results of our study therefore support \cite{pottasch06} conclusion
that the argon abundance is essentially the same in young objects (B stars, Orion nebula) and
in older PNe in the solar neighborhood.

\subsection{ISM absorption studies}

Because of its high ionization potential, argon is mostly in the neutral state
in the ISM. Argon abundances can be derived from column densities
measured from the far-ultraviolet resonance lines, \ion{Ar}{1} $\lambda\lambda 1048, 1067$.
\cite{sofia98} detected these \ion{Ar}{1} lines towards several early-type stars using
the interstellar medium absorption profile spectrograph during the first {\sl ORFEUS-SPAS\/}
shuttle mission in 1993 September. They found deficiencies in \ion{Ar}{1}/\ion{H}{1}
ranging from 0.2 to 0.6 dex, relative to the Ar/H abundance ratio in the Sun and in B stars
\citep{SUN89, keenan90}. They argued that this result does not provide an
indication of argon depletion onto dust grains, but it rather reflects selective argon
ionization relative to hydrogen because of the larger \ion{Ar}{1} photoionization
cross-section.

With the launch of the {\sl Far Ultraviolet Spectroscopic Explorer\/}, many
more line-of-sights in the ISM have been observed.
\cite{lehner03} carried out an analysis of 23 line-of-sights
toward white dwarfs and derived argon abundances in the local ISM. They confirmed 
deficiencies of the order of 0.4~dex, as found earlier by \cite{sofia98}. Moreover, 
the idea that these deficiencies actually map the local ISM ionization is well supported
by an increase of the abundance scatter outside the local Bubble.
It is therefore essential to have a
good reference value of the local argon abundance for ISM ionization studies, but 
\ion{Ar}{1} column densities are thus of limited use for accurately determining
the reference argon abundance in the solar neighborhood.

\subsection{Coronal vs. photospheric abundances}

The standard solar argon abundance recommended by \cite{AGS05},
A(Ar)= 6.18 $\pm$ 0.08, has been derived
from measured abundance ratios in the solar corona and solar energetic particles.
However, significant differences in the chemical composition of the solar corona relative
to the photosphere were noticed very early on. These differences were found to correlate 
with the first ionization potential of the studied elements, the so-called ``FIP effect''.
While there is still no agreed theoretical explanation of the FIP effect, the current general
consensus is that elements with lower FIP ($\la 10$\,eV) are enhanced in the corona
by a factor 4 to 6, while elements with higher FIP ($\ga 10$\,eV) have essentially
photospheric abundances \citep{favata03}.  However, most coronal studies based
for instance on \ion{Ar}{14}~$\lambda 4412$ \citep[e.g., ][]{young97}
or on helium-like argon lines observed in X-ray flares \citep{veck81, phillips03}
indicate an argon abundance, typically in a range between $6.2\la$~A(Ar)~$\la 6.4$,
that is lower than in Orion stars and nebula. The Orion results thus support the idea
that the solar corona exhibits significant underabundances of the two elements
with a high FIP ($\ga 15$\,eV). However, we cannot define the origin of these underabundances.
The FIP effect might either cover
a whole range, from overabundances for low FIP elements, to photospheric abundances
at intermediate FIP and underabundances for elements with the higest FIP, or
different physical processes might be in play and the behavior of low FIP and high FIP
elements might be unrelated. 

Since the neon and argon emission lines observed in the coronal and
chromospheric spectra of the Sun are potentially affected by the FIP effect,
\cite{feldman03} argued that it would be much preferable to observe and analyze
unmodified photospheric plasma. During a low-altitude impulsive flare,
photospheric plasma raised to flare temperature was observed by {\sl Skylab\/} \citep{feldman90}.
From the high measured plasma density and a Mg/O abundance ratio similar to
the photospheric value, Feldman~\& Widing established the origin of the emission.
They determined the Ar/Mg and Ne/Mg abundance ratios, yielding A(Ar) = 6.57 $\pm$ 0.12 and
A(Ne) = 8.08 $\pm$ 0.10,
with A(Mg) = 7.53 from \cite{AGS05}. These values are in excellent agreement
with the argon and neon abundances measured in B stars and in the Orion nebula.

Argon abundances can also be measured in the corona of other cool stars where the FIP effect
might not be as prevalent as in the Sun. \cite{maggio07} recently reported on a large study
of X-ray bright pre-main sequence (PMS) stars in Orion. They derived an average argon abundance of
A(Ar) = 6.51$^{+0.26}_{-0.17}$. The larger scatter implies that this result remains consistent
with the abundance measured in Orion B-type stars and, therefore, 
does not support an underabundance of argon in the corona of Orion PMS stars similar to the
underabundance found in the solar corona.

\subsection{{\sl In-situ\/} measurements in the Solar system}

Argon abundances have also been measured from energetic particles in the solar
wind, from lunar soils, and meteorites. \cite{reames98} contended that the most
comprehensive measurements of element abundances in the solar corona comes from
measurements of energetic particles that have been ejected during coronal mass
ejections. Because of their origin, we expect however that these abundances to show the
FIP effect. \cite{reames98} reported an abundance ratio Ar/O = 0.0033~$\pm$~0.0002.
Using \cite{AGS05} oxygen abundance, this ratio translates to A(Ar) = 6.18 $\pm$ 0.03.
\cite{cerutti74} reported a similarly low Ar/O ratio in lunar soils, yielding
A(Ar) = 6.26 $\pm$ 0.1. Finally, C1 chondrites reveal strong depletion because of
the volatile nature of noble gases. For instance, \cite{AGS05} list an argon abundance
in meteorites that is 6 to 7 orders of magnitude lower than all other measurements.

\subsection{Abundances in the Solar neighborhood}

Table~\ref{tab:sites} summarizes the argon abundances measured in different sites
in the Solar system and in the solar neighborhood, using a broad variety of methods.
Excluding meteorites, argon abundances span a range of 0.6~dex (a factor of 4), 
6.1~$\la$~A(Ar)~$\la 6.7$. On one hand, solar corona and {\sl in-situ} measurements
in the Solar system yield a low value (6.2~$\la$~A(Ar)~$\la 6.4$, or A(Ar) = 6.18 as
adopted by Asplund et~al.). On the other hand, Orion B-type stars and nebula provide
a tight agreement with emerging photospheric material during an impulse flare,
6.57~$\la$~A(Ar)~$\la 6.66$. What is therefore the best value to represent the argon
abundance in the Solar system? Can we use Orion
stars and nebula as proxies to the Sun in order to determine standard abundances
in the solar neighborhood?

Table~\ref{tab:solab} lists the abundances of light elements in the Sun, Orion B stars,
and Orion nebula, showing an excellent agreement for all elements if we adopt
\cite{feldman03} values for neon and argon in the solar photosphere. This table has
been constructed from representative studies that we view as current state-of-the-art works
in different objects, and without intending to review comprehensively all abundance studies
in the solar neighborhood (in particular for C, N, and O). We aim at putting
the present study in a broader context and discuss its relevance to the Ar solar abundance question.
First, we need to emphasize
that the agreement between the nebular and the photospheric analyses for all five 
elements (C, N, O, Ne, and Ar) cannot be fortuitous. It rather demonstrates the degree
of reliability that can be achieved today in abundance analyses. We believe that we may
thus reasonably claim that the accuracy of these abundances is 0.1~dex or better.
Second, the long-standing puzzle regarding the C, N, O abundances in the Sun and in
young OB stars appears now to have been resolved. In \cite{SUN89} reference work, the CNO abundances
were higher in the Sun than in OB stars, in an apparent contradiction with
the older age of the Sun. Recent studies
have now led to revise downward the solar abundances, and \cite{AGS05} recommend
C, N, O solar abundances that are very close to the values derived for Orion stars and
nebula. Therefore, these two arguments provide a robust support to the idea
that the composition of Orion's young material (in stars and in the nebula) is
very similar to the solar composition; hence, Orion's studies may be good proxies
to determine the composition in the solar neighborhood, in particular for neon and
argon too.

Let us assume on the other hand that the coronal abundances are representative of
the standard composition in the Solar system. This assumption thus implies that
the chemical enrichment over the last 5~Gyr would have been a factor of 2 for neon,
and a factor of 2 to 3 for argon, while C, N, and O remained virtually unchanged.
Analyses of PNe do not support such an enrichment of neon and argon over the last
few billion years. Within 1~kpc of the Sun, 7 PNe have mean abundances of
$<$A(O)$>$~=~8.66, $<$A(Ne)$>$~=~8.12 and $<$A(Ar)$>$~=~6.50 \citep{pottasch06},
in good agreement with Orion's results (see Table~\ref{tab:solab}). The 7 PNe
reveal however an enrichment in carbon and nitrogen resulting from nucleosynthesis
in the late evolution of their parent stars.
We therefore conclude that neon and argon do not reveal a strong chemical enrichment
in the solar neighborhood during the last few billion years, and that the 
depletion of neon and argon in the solar corona most likely arise from
yet little-understood processes in the corona, such as the FIP effect.

% -----------------------------------------------------------------------------

\section{CONCLUSIONS}

We have completed a new analysis of \ion{Ar}{2} lines in the blue spectrum
of 10 B main-sequence stars in the Orion association. The analysis involved
a fully consistent NLTE treatment based on our NLTE line blanketed model
atmospheres. We have derived a mean argon abundance which is in excellent agreement
with recent results for the Orion nebula \citep{esteban04}. However, the derived
abundance is significantly higher (0.45~dex) than the standard solar value recommended in 
the recent compilation of \cite{AGS05}. Based on a review of available argon
abundance determinations in various sites in the solar neighborhood, we argue
that the Orion's results much more likely represent the abundance
of argon (and neon) in the solar neighborhood, while the lower coronal abundances
reflect a depletion due to physical processes in the corona such as the FIP effect. 

Therefore, we propose to adopt the following values as new reference abundances in
the neighborhood of the Sun in the Galactic disk:
\begin{center}
A(Ne)~=~8.09~$\pm$~0.06 \\
A(Ar)~=~6.63~$\pm$~0.10 \\
\end{center}
The argon abundance is in excellent agreement with Lodders (2008) recent work,
in which she adopts for the solar system (protosolar) abundance, A(Ar)~=~6.57~$\pm$~0.10.
These abundances yield the abundance ratios Ne/O~=~0.26 and Ar/O~=~0.009.
We believe that these values should provide valuable reference points
for various fields of investigations, from the solar corona to the ionization of
the interstellar medium.

\acknowledgements
We thank Martin Asplund for inquiring about the argon abundances in B stars,
Carrie Trundle for sending equivalent widths of \ion{Ar}{2} lines measured in the
star HD~135485 and Letizia Stanghellini for discussions.
The work reported here is supported in part by the National Science
Foundation through AST03-07534, AST03-07532, AST06-46790 and NASA through NAG5-9213.

% ------------------------------------------------------------------------------------

% -----------------------------------------------------------------------------

\clearpage
\begin{deluxetable}{lllll}
%\tabletypesize{\small}
\tablewidth{0pt}
\tablecaption{Observing log.  \label{tab:exp}}
\tablehead{
\colhead{Star ID} & \colhead{Spectral Type} & \colhead{$V$} &
\colhead{UT Date} &  \colhead{Exposure times [s]}
}
\startdata
HD 35039  & B2 IV-V & 4.70     & 2007-01-30 & 2x150  \\
HD 35299  & B1.5 V  & 5.68     & 2007-01-30 & 2x300  \\
HD 35912  & B2 V    & 6.38     & 2007-02-06 & 720  \\
\nodata   & \nodata & \nodata  & 2007-03-03 & 2x700  \\
HD 36285  & B2 IV-V & 6.32     & 2007-03-03 & 2x600 \\
HD 36351  & B1.5 V  & 5.46     & 2007-02-06 & 330  \\
\nodata   & \nodata & \nodata  & 2007-03-03 & 500, 2x700  \\
HD 36591  & B1 IV   & 5.33     & 2007-02-06 & 180  \\
HD 36959  & B1 V    & 5.67     & 2007-02-06 & 2x270  \\
HD 37209  & B1 V    & 5.72     & 2007-02-06 & 360  \\
HD 37356  & B2 IV-V & 6.18     & 2007-03-03 & 2x600 \\
HD 37744  & B1.5 V  & 6.20     & 2007-02-06 & 480  
\enddata
\end{deluxetable}

% -----------------------------------------------------------------------------

\clearpage
\begin{deluxetable}{crrc}
%\tabletypesize{\small}
\tablewidth{0pt}
\tablecaption{Selected \ion{Ar}{2} lines and oscillator strengths.  \label{tab:line}}
\tablehead{
\colhead{Wavelength [\AA]}  & \colhead{$E_{\rm exc}$ [cm$^{-1}$]} & \colhead{$\log gf$} & 
\colhead{Accuracy\tablenotemark{a}}
}
\startdata
4426.001 & 135\,085.9960Ê&  0.158 & A \\
4430.189 & 135\,601.7336 & -0.174 & A \\
4589.898 & 148\,620.1411 &  0.100 & A \\
4657.901 & 138\,243.6442 & -0.236 & B \\
4726.868 & 138\,243.6442 & -0.103 & A \\
4735.906 & 134\,241.7392 & -0.108 & A \\
4764.865 & 139\,258.3384 & -0.06\phantom{0} & B \\
4806.020 & 134\,241.7392 &  0.210 & A \\
4847.810 & 135\,085.9960 & -0.223 & B \\
4879.864 & 138\,243.6442 &  0.246 & A \\
5062.037 & 135\,601.7336 & -0.465 & B \\
\enddata
\tablenotetext{a}{Estimated accuracy is better than 3\% (A) or better than 10\% (B).}
\end{deluxetable}

% -----------------------------------------------------------------------------

\clearpage

\begin{deluxetable}{lcccc}
\tablecaption{Argon and oxygen abundances. \label{tab:resul}}
\tablewidth{0pt}
\tablehead{
\colhead{Star} & \colhead{$T_{\rm eff}$} & \colhead{$\log g$} & \colhead{A(Ar)} & \colhead{A(O)}}
\startdata
HD 35039 & 20550 & 3.74 & 6.62 $\pm$ 0.06 & 8.60 \\
HD 35299 & 24000 & 4.25 & 6.70 $\pm$ 0.04 & 8.57 \\
HD 35912 & 19590 & 4.20 & 6.66 $\pm$ 0.05 & 8.70 \\
HD 36285 & 21930 & 4.40 & 6.59 $\pm$ 0.04 & 8.80 \\
HD 36351 & 21950 & 4.15 & 6.59 $\pm$ 0.05 & 8.76 \\
HD 36591 & 26330 & 4.20 & 6.73 $\pm$ 0.09 & 8.60 \\
HD 36959 & 24890 & 4.40 & 6.67 $\pm$ 0.07 & 8.76 \\
HD 37209 & 24050 & 4.15 & 6.63 $\pm$ 0.04 & 8.83 \\
HD 37356 & 22370 & 4.15 & 6.76 $\pm$ 0.04 & 8.67 \\
HD 37744 & 24480 & 4.40 & 6.65 $\pm$ 0.05 & 8.63 \\
\enddata
\end{deluxetable}

% -----------------------------------------------------------------------------

\clearpage

\begin{deluxetable}{lll}
\tablecaption{Argon abundances in various sites. \label{tab:sites}}
\tablewidth{0pt}
\tablehead{
\colhead{A(Ar)} & \colhead{Sites, methods} & \colhead{References} }
\startdata
6.66 $\pm$ 0.06  & B stars; NLTE   &  This study \\
6.50 $\pm$ 0.05  & B stars; NLTE   &  \cite{holmgren90} \\
6.54 $\pm$ 0.15\tablenotemark{a} & B stars; LTE & \cite{keenan90} \\ [2mm]
6.62 $\pm$ 0.05  & Orion nebula, $t^2=0.022$   &  \cite{esteban04} \\
6.6~~--~~6.7       & Orion nebula, $t^2=0.0 - 0.35$   &  \cite{peimbert77} \\[2mm]
6.50 $^{+0.14}_{-0.22}$ & Planetary nebulae, $d\leq 1$\,kpc & \cite{pottasch06} \\ [2mm]
6.1\phantom{0} $\pm$ 0.2  & ISM, \ion{Ar}{1}~$\lambda\lambda1048, 1067$   & \cite{sofia98, lehner03} \\ [2mm]
6.62 $\pm$ 0.12  & Low-altitude impulsive flare, Ar/Mg & \cite{feldman90, feldman03} \\[2mm]
6.45 $\pm$ 0.03  & Solar corona, X-ray flares, \ion{Ar}{17}   &  \cite{phillips03} \\
6.23 $\pm$ 0.1  & Solar corona, \ion{Ar}{14}~$\lambda4412$   &  \cite{young97} \\
6.38 $^{+0.18}_{-0.30}$  & Solar corona, X-ray flares   & \cite{veck81} \\ [2mm]
6.51 $^{+0.26}_{-0.17}$  & PMS stars in Orion, corona, X-rays   & \cite{maggio07} \\ [2mm]
6.18 $\pm$ 0.03  & Solar energetic particles, Ar/O   &  \cite{reames98} \\
6.26 $\pm$ 0.1  & Lunar soils, Ar/O   &  \cite{cerutti74} \\
\enddata
\tablenotetext{a}{Corrected as discussed in text.}
\end{deluxetable}

% -----------------------------------------------------------------------------

\clearpage

\begin{deluxetable}{llllll }
\tablecaption{Abundances in the solar neighborhood. \label{tab:solab}}
\tablewidth{0pt}
\tablehead{
\colhead{Element} & \colhead{Sun} & \colhead{Orion B stars}  & \colhead{Orion nebula} &
\colhead{PNe ($d <$~1\,kpc)} & \colhead{References} }
\startdata
He & 10.98 $\pm$ 0.02 &   \nodata        & 10.988 $\pm$ 0.003 & \nodata & 2,..., 7, ... \\
C  &  8.39 $\pm$ 0.05 &  8.35 $\pm$ 0.05 &  8.42 $\pm$ 0.02 & 8.65$^{+0.16}_{-0.27}$ & 1, 4, 7, 8 \\
N  &  7.78 $\pm$ 0.06 &  7.78 $\pm$ 0.07 &  7.73 $\pm$ 0.09 & 8.13$^{+0.14}_{-0.22}$ & 1, 4, 7, 8 \\
O  &  8.66 $\pm$ 0.05 &  8.70 $\pm$ 0.09 &  8.65 $\pm$ 0.03 & 8.66$^{+0.06}_{-0.07}$ & 1, 4, 7, 8 \\
Ne &  8.08 $\pm$ 0.10 &  8.11 $\pm$ 0.04 &  8.16 $\pm$ 0.09 & 8.12$^{+0.14}_{-0.20}$ & 3, 5, 7, 8 \\
Ar &  6.57 $\pm$ 0.12 &  6.66 $\pm$ 0.06 &  6.62 $\pm$ 0.05 & 6.50$^{+0.14}_{-0.22}$ & 3, 6, 7, 8
\enddata
\tablerefs{(1)~\cite{AGS05}; (2)~\cite{christensen98};
   (3)~\cite{feldman03};
   (4)~\cite{cunha94}; (5)~\cite{cunha06};
   (6)~This paper; (7)~\cite{esteban04}; (8)~\cite{pottasch06}. }
\end{deluxetable}

% -----------------------------------------------------------------------------
%%
%%        FIGURES
%%
% -----------------------------------------------------------------------------

\clearpage

\clearpage

\clearpage

% -----------------------------------------------------------------------------

\end{document}